\documentclass[a4paper,aip,apl,amsmath,amssymb,reprint,superscriptaddress]{revtex4-1}
\usepackage{graphicx}
\usepackage{amsmath}
\usepackage{dcolumn}
\usepackage{bm}
\usepackage{bbold}
\usepackage{color}
\usepackage{amsfonts}
\usepackage{amssymb}
\usepackage{mathrsfs}
\usepackage{tabularx}
\usepackage{braket}
\usepackage{mathtools}
\usepackage{soul}			

\usepackage{caption}
\usepackage{subcaption}

\renewcommand\Re{\operatorname{\mathfrak{Re}}}
\renewcommand\Im{\operatorname{\mathfrak{Im}}}

\begin{document} 
\title{Probing Dark Universe with Exceptional Points}
\author{Maxim Goryachev}
\affiliation{ARC Centre of Excellence for Engineered Quantum Systems, School of Physics, University of Western Australia, 35 Stirling Highway, Crawley WA 6009, Australia}

\author{Ben McAllister}
\affiliation{ARC Centre of Excellence for Engineered Quantum Systems, School of Physics, University of Western Australia, 35 Stirling Highway, Crawley WA 6009, Australia}

\author{Michael E. Tobar}
\email{michael.tobar@uwa.edu.au}
\affiliation{ARC Centre of Excellence for Engineered Quantum Systems, School of Physics, University of Western Australia, 35 Stirling Highway, Crawley WA 6009, Australia}

\date{\today}


\begin{abstract}

It is demonstrated that detection of putative particles such as paraphotons and axions constituting the dark sector of the universe can be reduced to detection of extremely weak links or couplings between cavities and modes. This method allows utilisation of extremely sensitive frequency metrology methods that are not limited by traditional requirements on ultra low temperatures, strong magnetic fields and sophisticated superconducting technology. We show that exceptional points in the eigenmode structure of coupled modes may be used to boost the sensitivity of dark matter mediated weak links. We find observables that are proportional to fractional powers of fundamental coupling constants. Particularly, in the case of axion detection, it is demonstrated that resonance frequency scaling with $\sim \sqrt{g_{a\gamma\gamma}\theta}$ and $\sim \sqrt[3]{g_{a\gamma\gamma}\theta}$ dependencies can be realised in a ternary photonic cavity system, which is beneficial as these coupling constants are extremely small. 

\end{abstract}

\maketitle

\section*{Introduction}

Many hypothesised candidates for dark matter are proposed to have weak coupling to established standard model particles, such as photons\cite{Goodsell:2009aa,Abel:2008aa}. This includes axions\cite{Peccei:2008aa,Preskill:1983aa} and paraphotons\cite{Okun,Jaeckel:2008aa} which have both been subjects of extensive experimental searches in the last few decades\cite{Cameron:1993aa,Chou:2008aa,Afanasev:2009aa,Pugnat:2014aa,Robilliard:2007aa,Asztalos:2010aa}, and recently there has been a renewed interest with many new and forthcoming experiments\cite{Parker:2013aa,McAllister:2017aa,CAPPToroid,HAYSTAC}. Traditionally, the detection methods are based on low noise power detection techniques, where one searches for excess photon power created by the interaction process between photons and these putative particles\cite{Asztalos:2010aa}. Much progress has been made in the design and optimisation of such experiments, with particular recent focus on the push to high ($>1$ GHz) and low ($<100$ MHz) frequency ranges~\cite{MADMAX,ABRACADABRA,ADMXDesign,XSWisp,CAPPPhase,CAPPPIZZA,RCHalo,BEAST,DielectricRing,AxionArray}. Such particles may come from some extra-terrestrial sources such as galactic halo axions, or be created artificially with another interaction process, such as Light Shining Through A Wall (LSW) experiments\cite{Povey:2010aa,Wagner:2010aa}. 

A conceptually different approach is based on detecting frequency splittings in the spectrum of a two-fold degenerate system, which can be broken due to the existence of a weak link between the two photonic cavities mediated by an intermediate interaction with a hypothetical particle\cite{Parker:2013aa}. In this approach, hypothetical particles created in one cavity are transferred into another, interacting with the second cavity photons and vice a versa creating a weak link that shifts eigenfrequencies of the system. Such a method of detection measures the lifting of the degeneracy of the system as a frequency shift, a technique commonly used in physics and engineering. Indeed, with the emergence of ultra-high Quality factor cavities in the optical and microwave regions of the photon spectrum, the frequency shift approach has found numerous applications, including biosensing, magnetic field and mass sensing, etc\cite{Frank:2012aa,Du:2017aa,Foreman:2015aa} and has been proposed as a means to detect paraphotons previously\cite{Parker:2013aa}. Many well developed frequency measurement techniques have been devised, which allow for high sensitivity of these kinds of measurements when applied to particular sensing problems. Similar frequency control methods of frequency comparison of uncoupled oscillators, frequency generators, are used in rotating experiments to search for Lorentz violating modifications to the standard model extensions\cite{Nagel:2015aa,Lo:2016aa}. { More generally, frequency metrology has established itself as a leading tool in the search for new physics\cite{Rosenband:2008aa, Derevianko:2016aa, Guerlin:2015aa,freqmetrology}.}

Unlike in the power detection scheme where the detection limit is set by noise sources in the measurement chain and thus ultimately limited by fundamental stochastic processes, the frequency detection approach is limited by mode linewidths and source frequency stability, that have been pushed by many orders of magnitude with ongoing progress in these areas. For example, ultra-narrow linewidth systems have achieved Quality factors of $>10^9$ for microwave and optical resonators\cite{Creedon:2012aa,DelHaye:2013aa}. Due to these achievements, frequency stability of microwave cryogenic sapphire clocks is approaching $10^{-16}$ level\cite{Hartnett:2006aa, Fluhr:2016aa}, whereas optical clocks have reached $10^{-18}$ level of stability and accuracy\cite{Bloom:2014aa}. In the same time, uncertainty of modern voltage standards based on Josephson junctions is only parts in $10^{-8}$ level\cite{Khorshev:2016aa}. This comparison suggests that the frequency metrology approach might be generally a more sensitive tool in detecting physics of the Dark Universe.


One way to further improve the sensitivity of a twofold-degenerate resonant detector is to move from a system degenerate only in eigenfrequencies and represented by a Diabolic Point (DP), to a system also degenerate in eigenstates and represented by an Exceptional Point\cite{Wiersig:2014aa,Wiersig:2016aa}. Existence of these points in non-Hermitian systems has been analysed theoretically\cite{PhysRevA.90.013802,PhysRevE.69.056216} and observed experimentally in different physical contexts\cite{Hodaei:2015aa,Hodaei:2017aa,PhysRevLett.86.787,PhysRevLett.103.134101,Dietz:2007aa}. It has been demonstrated that for a sufficiently small perturbation $\frac{\varepsilon}{\omega_0}\ll1$ acting on one of {the} resonant subsystems with angular frequency $\omega_0$, the corresponding observable frequency shift of the total system is proportional to $\varepsilon^{1/2}$ if the detector works at the doubly degenerate EP. This result is a clear advantage over an $\varepsilon$ dependent frequency shift observed in the case of a DP. Moreover, the power of the shift inversely scales with the order of degeneracy of the system, $n$, giving in general a $\varepsilon^{1/n}$ dependence\cite{Hodaei:2017aa}. In this work, we consider applications of this idea to detection of weak links arising in the context of detection of theoretical fundamental particles. In this problem the perturbation $\epsilon$ acting on one or both subsystems, for example, cavities, is replaced with a weak link ${\zeta}\ll{\omega_0}$ between them.

\section{Weak Links from Dark Universe}

In the following sections, we will determine and compare sensitivities to weak links $\zeta$ of different detectors by analysing their dynamics in the slowly varying envelope approximation where it is governed by the Schr{\"o}dinger-type equation ($\hbar=1$)\cite{Wiersig:2016aa,Siegman:1986aa}:
\begin{multline}
\begin{aligned}
	\label{dynamics}
	\displaystyle  \frac{d}{dt}\psi = -iH(\zeta)\psi,
\end{aligned}
\end{multline}
{where $\psi$ is a wave function, $\zeta$ is the weak link parameter, i.e. the coupling strength between two modes or cavities in frequency units, and $H$ is a non-Hermitian matrix representing envelope dynamics of the system in a fast rotating frame. We refer to this matrix $H$ as an effective Hamiltonian.} In this section, we consider how such weak links $\zeta$ may appear from two hypothetical dark sector particles, namely paraphotons and axions. 

\subsection{Paraphotons}

Paraphotons are Weakly Interacting Slim Particles with sub-eV masses that couple to photons via kinetic mixing. Laboratory searches for these hidden sector particles put bounds on the strength of the kinetic mixing and are dominated by {LSW} type experiments\cite{PhysRevD.82.052003,Wagner:2010ab}. In this experiment two spatial regions, e.g. cavities, are separated with a wall impenetrable to normal photons. One of these regions is strongly pumped with photons, which generates paraphotons that can penetrate the other region and mix with the thermal and vacuum photons in that cavity. This effect can be detected as an excess power in the second region. It should be noted that this approach is not sensitive to galactic halo dark matter paraphotons, but to paraphotons generated inside the first region. The sensitivity of these approaches are limited not only by the noise floor of the setup but more by systematic effects such as direct leakage of photons into the detection region. Also, some constraints on paraphotons can be obtained from cosmic microwave background data\cite{Mirizzi:2009aa} and Coulomb
law experiments\cite{Bartlett:1988aa}.  Alternatively, for the {LSW} type experiments, one may search for modifications of the system eigenvalues as a response to the paraphoton mediated cavity coupling\cite{Parker:2013aa}. In this section we demonstrate the appearance of this coupling.

The Equations of Motion (EOM) for two cavities with photon fields $C_i(\mathbf{x},t)$ and paraphoton field $B(\mathbf{x},t)$ from the paraphoton-photon Lagrangian\cite{Jaeckel:2008aa} are given by;
\begin{multline}
\begin{aligned}
	\label{para1}
	\displaystyle  
 \left( \begin{array}{ccc}
\partial^\mu\partial_\nu + \xi^2 m^2_{\gamma\prime}  & 0 & -\xi m^2_{\gamma\prime} \\
0 & \partial^\mu\partial_\nu + \xi^2 m^2_{\gamma\prime} & -\xi m^2_{\gamma\prime} \\
-\xi m^2_{\gamma\prime} & -\xi m^2_{\gamma\prime}  & \partial^\mu\partial_\nu + m^2_{\gamma\prime}\end{array} \right)
A = 0
\end{aligned}
\end{multline}
where $A = [C_1(\mathbf{x},t),C_2(\mathbf{x},t),B(\mathbf{x},t)]^T$, $\xi$ is the paraphoton-photon mixing parameter and $m_{\gamma\prime}$ is the paraphoton mass. 

In order to remove the spatial dependence of the fields in Eq.~\ref{para1}, one breaks the cavity fields into spatial and time dependent multipliers and introduces retarded massive Greens functions for the paraphoton field\cite{Parker:2013aa}. Moreover, on the time and space scale of any laboratory experiment the paraphoton field has an infinite nature. Particularly, due to negligible damping rate, the paraphoton dynamics can be neglected in the rotating frame associated with the cavity resonance frequencies. { Based on these assumption and integrating spacial distributions of the photonic modes, the system dynamics~(\ref{para1}) can be reduced to a two mode system with the following coupling strength\cite{Parker:2013aa}}
\begin{multline}
\begin{aligned}
	\label{coupling}
	\displaystyle  \zeta_\text{paraphoton} = -\frac{\xi^2m^4_{\gamma\prime}}{2\omega_0}\cdot \\
	\displaystyle \int_{V_1}d^3\mathbf{x}\int_{V_2}d^3\mathbf{y}\frac{\exp{ik_p|\mathbf{x}-\mathbf{y}|}}{4\pi|\mathbf{x}-\mathbf{y}|} C_1({\mathbf{x})\cdot C_2({\mathbf{y}}}),
	\end{aligned}
\end{multline}
where $V_i$ is the volume of the $i$th cavity and $\omega_0$ is the bare cavity resonance frequency. In principle, the resonance frequency of each cavity is also shifted due to coupling to paraphotons, but this effect can be neglected and cannot be observed in the present scheme as it cannot be modified without changing the cavity geometry. Overall, the resulting weak coupling is a function of the kinetic mixing and the paraphoton mass as well as cavity geometry. 

\subsection{Axions}

The mainstream approach to detect galactic halo axions\cite{Abbott:1983aa,Dine:1983aa} (a haloscope) employs the Sikivie-like detector aproach\cite{Sikivie:1983aa}. The detector utilises the inverse-Primakoff effect, in an interaction between virtual photons from a strong DC magnetic field and galactic halo axions, creating a real photon of a frequency corresponding to the axion mass. The resulting signal photons are detected using a resonant cavity as an antenna tuned to this frequency\cite{Asztalos:2010aa,McAllister:2017aa}. In principle, instead of a virtual photon supplied by a DC magnetic field, one can use real photons by engaging another cavity or a mode\cite{RfSikivie} and utilise the same power detection techniques. Contrary to this, one may also measure how the resonant frequencies of the two modes are shifted as a result of interaction between the two photons and an axion signal - this is the weak link problem. Indeed, it has recently been demonstrated that the weak coupling detection problem may be relvant to the search for galactic halo dark matter axions\cite{freqmetrology}. In this section, we briefly demonstrate this effect for the case of two modes that share the same volume. 

It is widely accepted that the axion modification to the standard electrodynamics\cite{Wilczek:1987aa} may be written as a Hamiltonian term of the form:
\begin{multline}
\begin{aligned}
	\label{axion1}
	\displaystyle  H_\text{axion} = \kappa \theta \int_V \mathbf{E}\cdot \mathbf{B} dv,
\end{aligned}
\end{multline}
where $\theta$ is a scalar axion-like field, $\mathbf{E}$ and $\mathbf{B}$ are standard electric and magnetic field vectors, and $\kappa$ is a coupling constant. In a case when two modes of resonance frequencies $\omega_1$ and $\omega_2$ share the same volume, one may rewrite the Hamiltonian in terms of corresponding creation (annihilation) operators $c_i^\dagger$ and $c_i$:
\begin{multline}
\begin{aligned}
	\label{axion2}
	\displaystyle  H_\text{axion}  = \kappa \theta \int_V (\mathbf{E}_1\cdot \mathbf{B}_2 + \mathbf{E}_2\cdot \mathbf{B}_1)dv=\\
	\displaystyle  i\hbar \frac{ g_{a\gamma\gamma}}{2}\sqrt{{\omega_1\omega_2}} \theta \Big[\xi_-(c_1c_2^\dagger-c_1^\dagger c_2) + \xi_+(c_1^\dagger c_2^\dagger-c_1 c_2)\Big]\\
\end{aligned}
\end{multline}
where $ g_{a\gamma\gamma}$ is axion-photon coupling strength, $\theta$ is the strength of the axion signal, and $\xi_\pm$ is the form factors characterising mutual orthogonality of the modes\cite{freqmetrology}. { This interaction Hamiltonian could be simplified by applying the Rotating Wave approximation, the process of elimination of rapidly-oscillating terms of interacting fields in the rotating frame. To perform this approximation, we need to assume a certain relationship between three fields. In this work we consider only the axion upconversion case where the frequency of axion induced photons equals the difference in frequencies between the modes $\omega_a = \omega_1-\omega_2$. By substituting $c_n\rightarrow c_n e^{-j\omega_n t}$, $a\rightarrow a e^{-j\omega_a t}$ and neglecting all fast oscillating terms, the coupling term may be represented as follows\cite{freqmetrology}:}
\begin{multline}
\begin{aligned}
	\label{axion4}
	\displaystyle H_\text{axion}/\hbar  = \zeta_\text{axion} (c_1 c_2^\dagger +  c_2 c_1^\dagger),
\end{aligned}
\end{multline}
which is a beam-splitter or swapping type of interaction. The value $\zeta_{axion}$ is proportional to the axion-photon coupling $\kappa$ and the amplitude of axion field $|\theta|$ as well as depending on geometry and frequency of both modes:
\begin{multline}
\begin{aligned}
	\label{axion5}
	\displaystyle \zeta_\text{axion} =  \frac{ g_{a\gamma\gamma}}{2}\sqrt{\frac{\omega_1\omega_2}{V_1V_2}} \theta \int_Vd^3r\big[\mathbf{e}_1\cdot \mathbf{b}_2-\mathbf{e}_2\cdot \mathbf{b}_1\big],
\end{aligned}
\end{multline}
where $\mathbf{e}_n(\mathbf{r})$ and $\mathbf{b}_n(\mathbf{r})$ are unit vectors representing the mode polarization. Thus, one can detect existence of galactic axions of frequency $\omega_a$ by measuring the weak link between two modes of different frequencies. We note that in this case, we are sensitive to galactic halo axion dark matter, as opposed to the paraphoton case where we are sensitive to generated paraphoton dark matter, as discussed above.

\section{Second Order EP Detection}

\begin{figure}
\includegraphics[width=0.6\columnwidth]{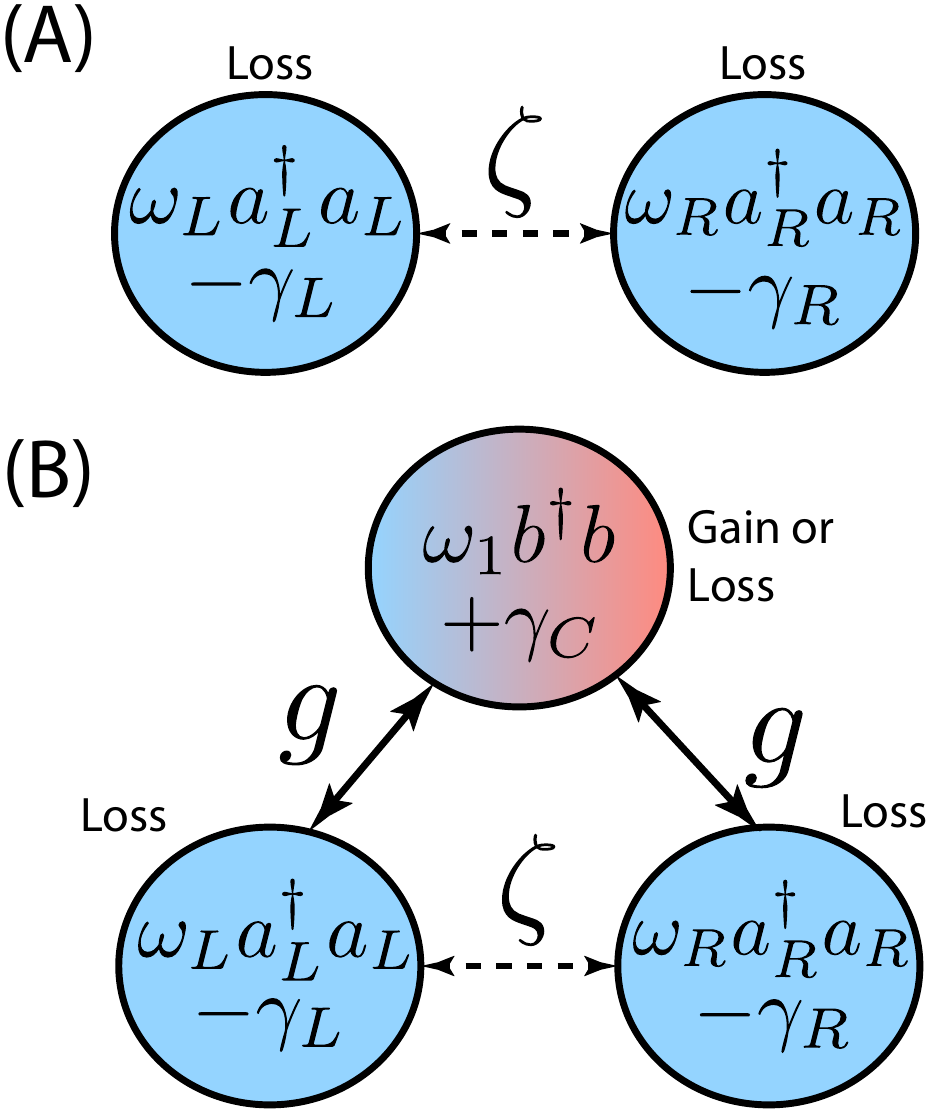}
\caption{Schemes for detecting weak links $\zeta$ between photonic { modes: (A) traditional scheme of frequency splitting detection between two lossy cavities using a DP, (B) ternary mode system that can be tuned to an EP.}}
\label{systems}
\end{figure}

To illustrate the improvement in sensitivity for the above techniques with EP detection, we consider two types of detectors. {The first detector is a traditional dual mode system exhibiting a DP with a weak link $\zeta$ between the modes  as depicted in Fig.~\ref{systems} (A). In the case of paraphoton detection, ideally both modes have the same angular frequency $\omega_0 = \omega_L=\omega_R$ and losses $\gamma_L=\gamma_R = \gamma$, whereas for galactic halo axion searches with such a system, the upconversion frequency relationship $\omega_a = \omega_L-\omega_R$ must be considered. Both cases lead to the same interaction hamiltonian in a corresponding rotating frame. }

\begin{figure}
\includegraphics[width=1\columnwidth]{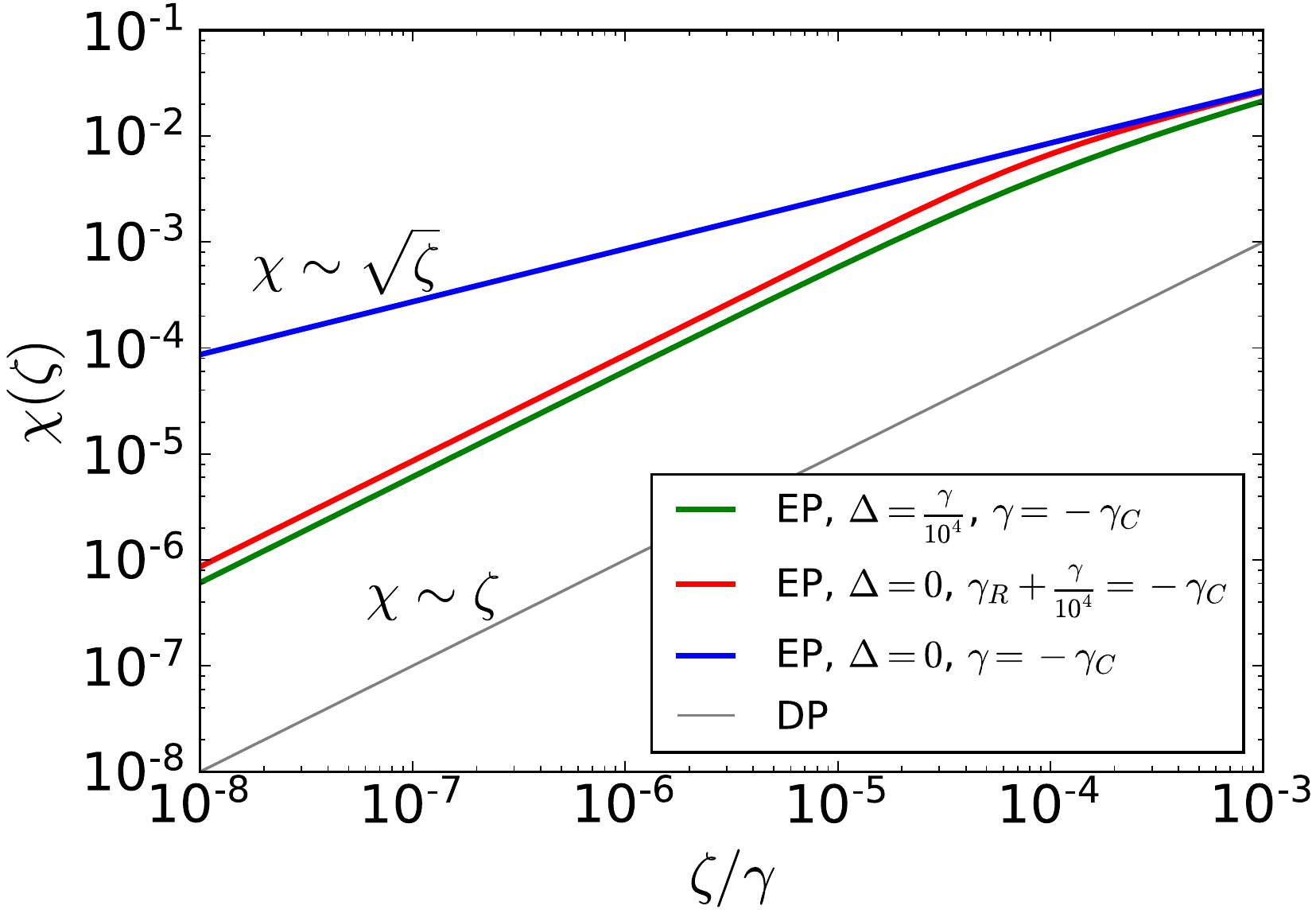}
\caption{Sensitivity of DP and EP detectors to a weak coupling $\zeta$. }
\label{sensitivity1}
\end{figure}

The second class of detector (the EP case) is a coupled mode system of three photonic modes with identical angular resonance frequencies $\omega_0$ (for the paraphoton case, and {the upconversion relationship} for the axion case as per above). Two side cavities denoted $L$ and $R$ exhibit loss characterised by the constant $\gamma = \gamma_L=\gamma_R$ whereas for the central cavity $C$  we consider both gain $\gamma_C>0$ or loss $\gamma_C<0$ regimes. This ternary configuration is depicted in Fig.~\ref{systems} (B). Both lossy side cavities are coupled to the central cavity by engineered links with coupling strength $g$. Note that in axion non-degenerate case, the coupling $g$ has to be parametrically modulated to account for the frequency difference between the side modes. The mutual coupling between the left and right modes comes from axion or paraphoton physics and constitutes the weak link $\zeta$ that has to be detected. The effective Hamiltonian for an eigenvalue problem in the appropriate rotating frame can be written in the following matrix form
\begin{multline}
\begin{aligned}
	\label{ham1}
	\displaystyle \frac{H}{\gamma} =
 \left( \begin{array}{ccc}
{\Delta}-i & \widetilde{g} & \widetilde{\zeta} \\
\widetilde{g} & i(2\alpha-1) & \widetilde{g} \\
\widetilde{\zeta} & \widetilde{g} & -i\end{array} \right)
\end{aligned}
\end{multline}
where $2\alpha=\gamma_C/\gamma+1$, $\widetilde{g} = g/\gamma$, $\widetilde{\zeta} = \zeta/\gamma$ are quantities scaled to the loss constant $\gamma$, {and $\Delta$ is a possible mismatch between the left and right cavities. In the absence of a weak coupling $\zeta$ and the cavity mismatch}, the eigenfrequencies of the problem are 
\begin{multline}
\begin{aligned}
	\label{eig1}
	\displaystyle \frac{e_\pm}{\gamma} = i(\alpha-1)\pm\sqrt{2\widetilde{g}^2-\alpha^2}, e_0 = -i\gamma.
\end{aligned}
\end{multline}
It can be demonstrated that two of these eigenvalues and corresponding eigenvectors coalesce when $\alpha = \sqrt{2} \widetilde{g}$ corresponding to the second order EP of the system and giving the $\delta e= e_+-e_-\sim\epsilon^{1/2}$ sensitivity of the eigenvalues to the perturbation $\epsilon$ of one of eigenfrequencies. To check if the same sensitivity applies to the weak link $\zeta$, we calculate sensitivity as a shift in eigenvalues in the presence of $\zeta$ to its value without the link in units of side cavity losses:
\begin{multline}
\begin{aligned}
	\label{sens}
	\displaystyle \chi(\zeta) = \Re\frac{e_\pm(\zeta) - e_\pm(0)}{\gamma}.
\end{aligned}
\end{multline}
This quantity for ($\alpha = 3/2$) in comparison with the same characteristic for the DP case is shown in Fig.~\ref{sensitivity1}. The result suggests that unlike in the DP case where $\delta e\sim \gamma\chi\sim \zeta$, the EP detector tends to result in more favourable $\delta e\sim \sqrt\zeta$ dependence. The result is also calculated for a small mismatch between two side cavities appearing as non-zero frequency mismatch $\Delta = 10^{-3}\gamma$ and non-equal dissipation rates $\gamma_L=\gamma_R+10^{-3}\gamma$. As the system deviates from the EP in both cases, the performance of the EP detector is degraded for small $\zeta$ but still overcomes the DP case.

\begin{figure}
\includegraphics[width=1\columnwidth]{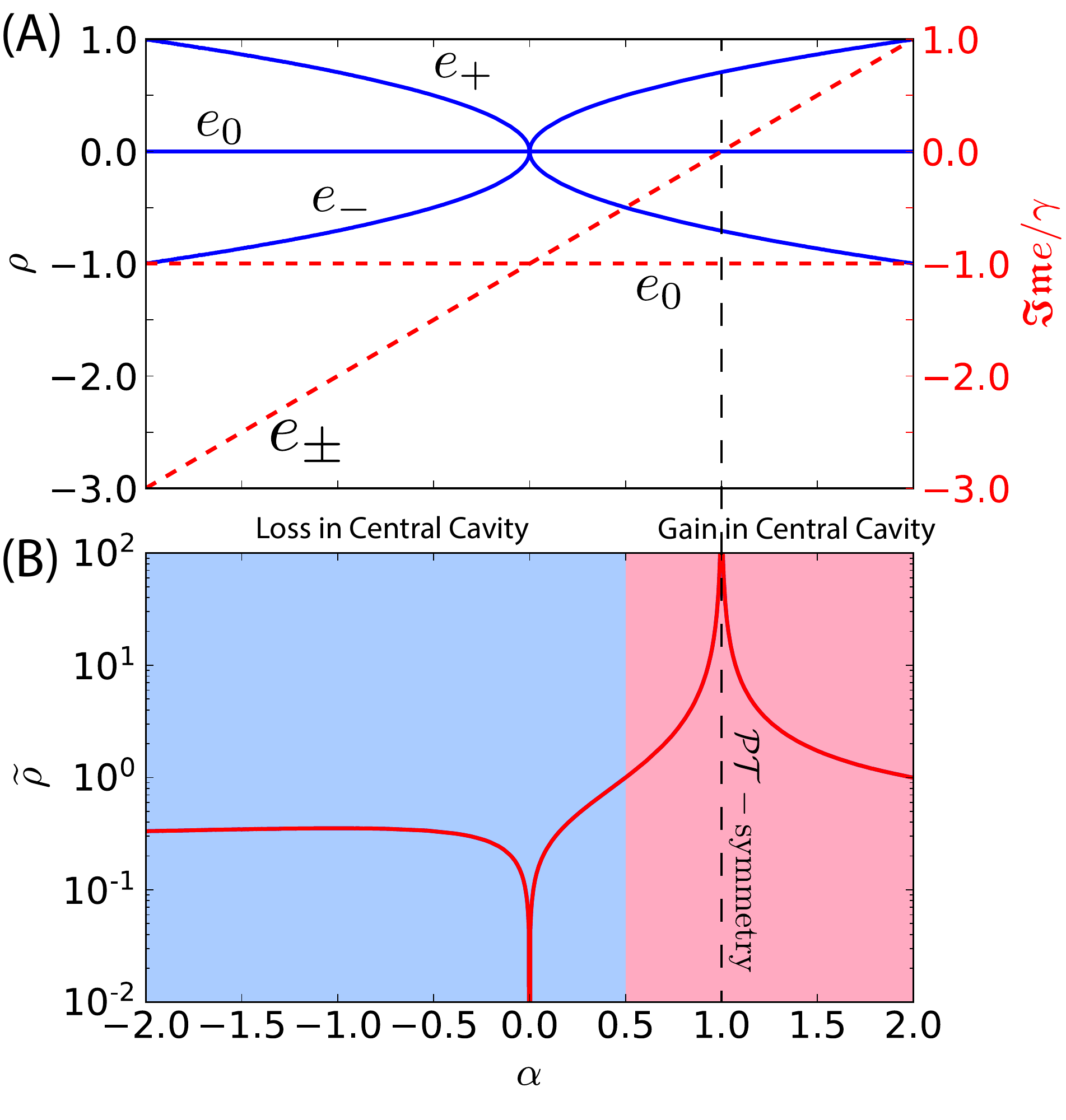}
\caption{(A) Dependence of the sensitivity coefficient $\rho_\pm$ and imaginary parts of eigenfrequencies on cavity mismatch $\alpha$. (B) Normalised sensitivity coefficient $\widetilde{\rho}_\pm$ showing the $\mathcal{PT}$-symmetric regime to be the most sensitive point. }
\label{PTtuning}
\end{figure}

With the EP condition setting the relation  $\alpha\gamma = \sqrt{2} {g}$, for given side cavity losses $\gamma$ and central cavity coupling $g$, the system still leaves a free parameter, the mismatch constant $\alpha$, which sets the central cavity loss/gain parameter $\gamma_C$. As described at the beginning of this section, this parameter can be either negative ($\alpha<\frac{1}{2}$), resulting in loss, or positive ($\alpha>\frac{1}{2}$), exhibiting net gain. 
In order to understand the role of this parameter, we numerically calculate the derivative of the real components of eigenfrequencies with respect to square root of coupling $\zeta$ at $\zeta=0$:
\begin{multline}
\begin{aligned}
	\label{sens2}
	\displaystyle \rho_\pm = \frac{\partial \Re e_\pm}{\partial \sqrt{\widetilde\zeta}}\Big|_{\widetilde\zeta = 0},
\end{aligned}
\end{multline}
which gives the proportionality constant of the $\delta\sim \rho_\pm\sqrt{\zeta}$ law for infinitesimally small values of $\zeta$. Dependence of this coefficient on the central cavity mismatch $\alpha$ is shown in Fig.~\ref{PTtuning} (A). The result suggests that despite the presence of a sign of $\alpha$, the magnitude of the scaling coefficient for $\delta$ increases with $|\alpha|$. At the same time the bandwidth of the resonances increases with negative $\alpha$ making the detection less efficient. 

In order to find the optimal point, we calculate the normalised version of the derivative
\begin{multline}
\begin{aligned}
	\label{sens3}
	\displaystyle \widetilde{\rho}_\pm = \frac{\rho}{\Im{e_\pm}},
\end{aligned}
\end{multline}
giving the coefficient in front of the frequency scaling law with respect to the resonance linewidths. The maximum of this value corresponds to the optimal detection point. The result is shown in Fig.~\ref{PTtuning} (B) where the peak value corresponds to $\alpha = 1$ or, the $\mathcal{PT}$-symmetric regime of the pair of resonances $e_\pm$. Indeed, as it is seen from Eq.~(\ref{eig1}), this value of mismatch nullifies the imaginary part of the two eigenvalues by equalising the overall losses and gain. For $\alpha \neq 1$, the $\mathcal{PT}$-symmetry is broken and either loss or gain dominates. 
So, it is important to note that {Fig.~(\ref{PTtuning}) (B)} demonstrates that the effect of $\sqrt{\zeta}$ scaling exists not only for the $\mathcal{PT}$-symmetric regime, where it is enhanced, but also for $\mathcal{PT}$-broken systems. Moreover, gain in the central cavity is not a necessary condition for the $\sqrt{\zeta}$ detection, which potentially simplifies possible physical realisations.  

{ Note that at $\alpha = 0$, the system cannot satisfy the EP condition as it would require the zero coupling to the central cavity. As a result the $\widetilde{\rho}_\pm$ drops to zero as demonstrated in Fig.~(\ref{PTtuning}) (B).}

\section{Higher Order EP Detectors}

\begin{figure}
\includegraphics[width=1\columnwidth]{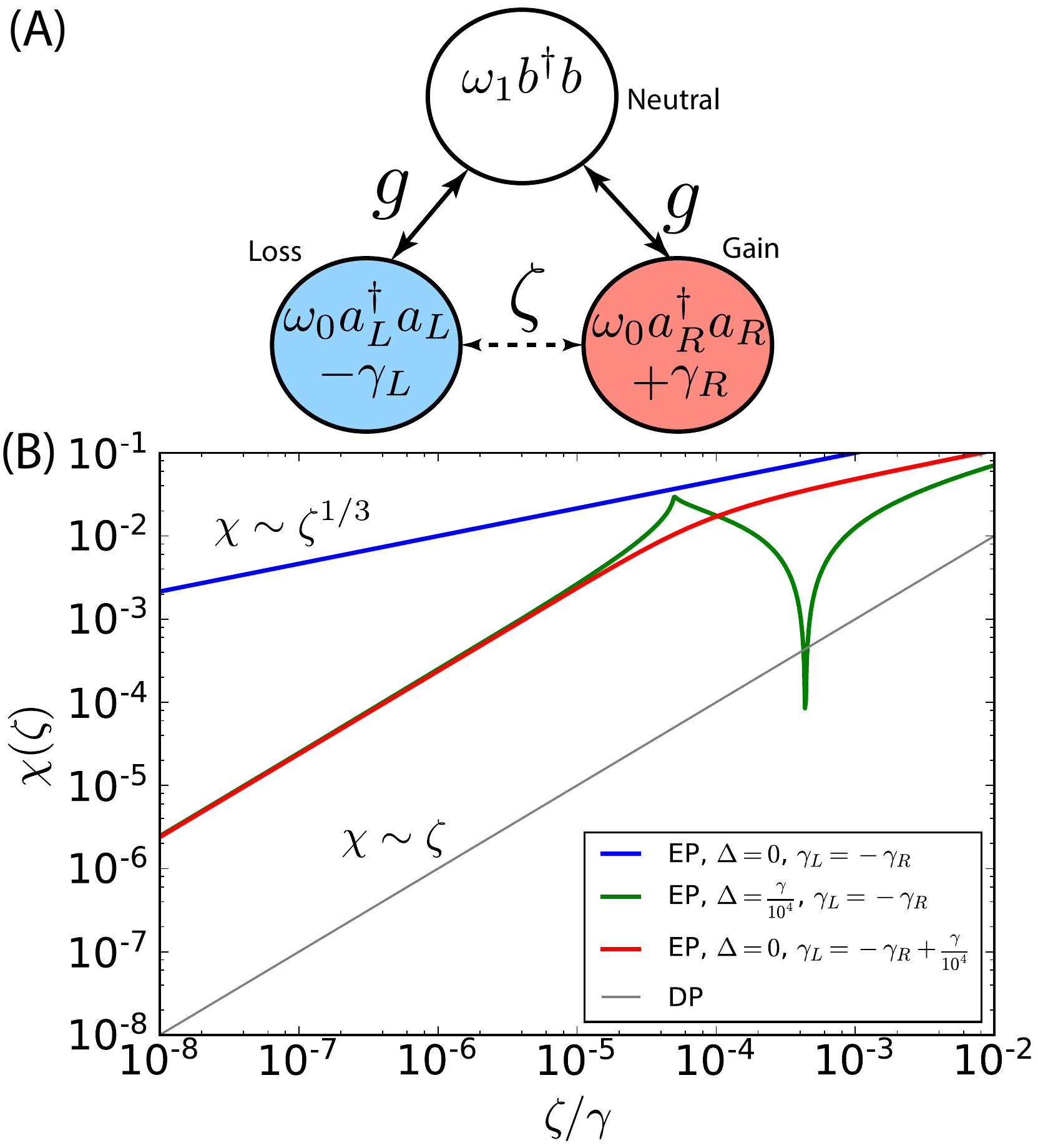}
\caption{(A) Detector exhibiting the third order EP. (B) Sensitivity of a third order EP detector to a weak coupling $\zeta$ in comparison to the DP case. }
\label{sensitivity2}
\end{figure}

With the addition of more cavities and links to the system, it is possible to construct higher order EPs resulting is scaling laws $\delta e  \sim \zeta^{1/N}$. Existence of these points has been recently demonstrated in a $\mathcal{PT}$-symmetric ternary system realised as a photonic laser molecule\cite{Hodaei:2017aa}. In this section, we analyse if such an arrangement may result in improved weak link detection. The ternary cavity structure consists of one lossy (left), one neutral (central) and one gain (right) cavity as shown in Fig.~\ref{sensitivity2} (A). Each of the left and right cavities is normally coupled to the central element with the strength $g$. The weak link to be detected is introduced between the right and left cavities. 

It can be easily demonstrated, that in the $\mathcal{PT}$-symmetry case ($\gamma_L=\gamma_R=\gamma$, $\gamma_C = 0$), the system exhibits a triple EP when $\gamma = \sqrt{2} g$. In this case all three eigenvalues as well as corresponding eigenvectors coalesce. The corresponding sensitivity to a weak link as defined in Eq.~(\ref{sens}) is shown in Fig.~\ref{sensitivity2} (B), which demonstrates the $\delta e \sim \zeta^{1/3}$ scaling law for the ideal EP detector as opposed to the DP case. Like in the case of the second order EP detector in the previous section, deviations from ideal conditions, such as non-zero cavity detuning $\Delta$ and mismatch in losses/gains, leads to degraded performance in the $\zeta\rightarrow 0$ limit. 

Additionally, since losses of the left cavity are completely compensated by the right one with the central cavity being neutral, the proposed detector works in the $\mathcal{PT}$-symmetry regime as manifested by the vanished imaginary component of the system eigenvalues. 

Regarding the overall sensitivity, the considered third order EP weak link detector would be advantageous over the one proposed in the previous section. On the other hand, it would be more challenging to realise in practise as it requires both a purely gain element and an element with fully compensated internal losses. 

\section{On Physical Realisation}

{\subsection{Frequency Metrology}

This work presents a general framework applied to dark matter detection based on frequency metrology and fractional sensitivity of dynamical systems working at EPs. As an extension of the frequency metrology approach, it is based on the same experimental techniques discussed in previous work\cite{Parker:2013aa,freqmetrology}. All these techniques boil down to a generation of signals with ultra low frequency fluctuations and the detection of small frequency or phase variations. This can be achieved in two ways\cite{freqmetrology}. Firstly, one can employ external low noise sources and measure eigenmode frequency deviations by comparing the source frequency with the frequency of the signal transmitted through or reflected from the system. Secondly, one may use eigenmodes of the system as a frequency selective element in a feedback oscillator and measure its frequency either in a dual loop or single loop approach\cite{freqmetrology}. Particular details of these experiments will depend on the targeted regions of the parameter space and compromise between complexity and sensitivity.

Comparing to existing frequency metrology techniques, the proposed approach provides orders of magnitude improvements in sensitivity due to more favourable scaling laws. To demonstrate this fact, we assume the widely accepted fact that the frequency can be effectively controlled and measured to about $10^{-6}$ of available cavity linewidth\cite{Rubiola:2008aa} that sets sensitivity for the given observation time. Thus a $10$GHz oscillator based on a dielectric resonator with $Q=10^8$ would be able to immediately detect induced variations on the order of $10^{-4}$Hz. In the same time, the second order EP based detector would be able to resolve variations on the order of $10^{-7}-10^{-8}$Hz.   
Although, this improvement is achieved due to increased complexity: the method requires precise engineering of gain/loss factors as a function of mode couplings. Exact implementation and adjustment of these details will be a base for future particular proposals.

Comparing the traditional power detection methods, the proposed approach relies on a completely different type of precision measurements. Instead of integrating over all available signal power with the lowest possible effective noise (either quantum or thermal), here we rely on precision control of external and internal uncertainties of oscillator or cavity parameters, such as temperature, vibration, nonlinearity, etc. The set of techniques that allow us to avoid these uncertainties constitutes the field of frequency metrology and control.  }

\subsection{Experimental Demonstrations}

The existence of Exceptional Points have been demonstrated both experimentally and theoretically in many physical contexts such as optically generated non-Hermitian photonic lattices\cite{Hahn:2016aa}, fiber-coupled whispering-gallery-mode (WGM) resonators\cite{Peng:2016aa, Hodaei:2017aa}, metamaterials\cite{Vaianella:2018aa}, quantum fluids\cite{Gao:2018aa}, electronic circuits\cite{Stehmann:2004aa}, acoustic systems\cite{Shi:2016aa}, etc. Thus, there is a wide range of physical platforms that can be used to realize the proposed dark matter detection schemes. Futhermore, small perturbation detection with EPs has been demonstrated previously in the optical frequency spectrum\cite{Hodaei:2017aa,Chen:2017aa}.  {Moreover, axion upconversion allows application of optical cavities and techniques for search of axions in microwave and RF frequency ranges\cite{freqmetrology}. The mainstream platform for these demonstrations as well as many other related $\mathcal{PT}$-symmetry experiments utilises WGMs in coupled microtoroids and an active medium (ion doped crystals) for gain. It should also be emphasised that unlike all traditional power detection techniques for axion detection, the frequency metrology approach does not require any external magnetic fields\cite{freqmetrology}, thus, allowing utilisation of magnetic impurities and magnetically active crystals as a source of gain.

Similar approaches may be taken in the microwave region of the spectrum: several gain materials have been investigated for application in masers\cite{Gerritsen:1962aa,Creedon:2010aa}, even for room temperature operation\cite{Breeze:2018aa}. Cryogenic masers and corresponding investigations of gain materials were a popular research field in 1960-70s giving a wide range of possible solutions to implement and control gain and loss parameters of microwave modes. Additionally, one may consider implementations using nonlinear discrete components such as diodes\cite{Ye:2014aa} or Josephson junctions\cite{Zorin:2017aa} that are allowed as DC magnetic fields are not required.
}

\subsection{Alternative Realisation of Gain}

Shifting from optical to microwave frequencies yields new opportunities for realisation of effective gain as one may employ elements such as linear external amplifiers. Indeed, by introducing an external amplifying feedback it is possible to reduce cavity bandwidth and at some level reverse the apparent sign of coupling to the environment. In this regime, a lossy cavity is turned into a signal generator.

\begin{figure}
\includegraphics[width=0.6\columnwidth]{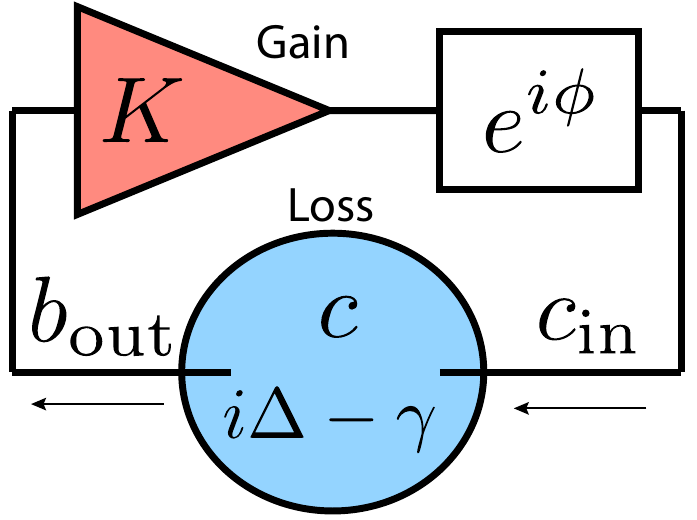}
\caption{Employing an external amplifier to control the cavity parameters.}
\label{systems2}
\end{figure}

In order to understand the idea of feedback compensation of cavity losses, we consider Equations of Motion (EOMs) for a cavity characterised with overall losses $\gamma$ and detuning $\Delta$ coupled to a unidirectional amplifier with gain $K$ and a phase shifter $\phi$ (see Fig.~\ref{systems2}):
\begin{multline}
\begin{aligned}
	\label{ampli}
	\displaystyle \frac{d}{dt}a = (i\Delta-\gamma) a -i\sqrt{2\gamma_b}b_\text{in} -i\sqrt{2\gamma_c}c_\text{in},\\
	\displaystyle c_\text{in} = Ke^{-i\phi}b_\text{out},\\
	\displaystyle b_\text{out}-b_\text{in} = -i\sqrt{2\gamma_b} a,\\
\end{aligned}
\end{multline}
where $a$, $b_\text{in}$ ($b_\text{out}$) and $c_\text{in}$ ($c_\text{out}$) are complex amplitudes for the cavity and two port input (output) fields. Here we neglect all noise terms including those introduced by the amplifier as the objective of the study is the eigenmode analysis. The resulting EOM can be written as:
\begin{multline}
\begin{aligned}
	\label{ampli2}
	\displaystyle \frac{d}{dt}a = (i\Delta-\gamma - 2\sqrt{\gamma_c\gamma_b}Ke^{-i\phi})a\\
	\displaystyle -i(\sqrt{2\gamma_b} +\sqrt{2\gamma_c}Ke^{i\phi})b_\text{in}.\\
\end{aligned}
\end{multline}
This EOM demonstrates that by changing the gain $K$ and phase shift $\phi$ that one is able to control both the real and imaginary parts of a corresponding diagonal element in the Hamiltonian (Eq.~\ref{dynamics}). In other words, one can achieve the overall gain regime for a lossy cavity with a positive feedback. 

\section{Conclusion}

The fundamental problem with detection of common dark matter candidates is the miniscule coupling to the observable sector of the universe. For many years, the fields of direct axion and paraphoton dark matter detection suffered from a scarcity of methods that are linear in the relevant small coupling constants\cite{freqmetrology,Stadnik:2014aa}. Only recently a few new techniques that are linear in coupling have been demonstrated. In this work we, for the first time, propose a concept in which an observable, resonance frequency shift is better than linear in coupling to the dark matter particle. In particular, we propose two ternary mode systems that can provide frequency shifts scaling as square and cubic roots of the small coupling constants. In the context of galactic halo axions this means frequency shifts proportional to $\sim \sqrt{g_{a\gamma\gamma}\theta}$ and $\sim \sqrt[3]{g_{a\gamma\gamma}\theta}$, greatly improving prospects for detection.


\section*{Acknowledgements }

This research was supported by the Australian Research Council Centre of Excellence for Engineered Quantum Systems (project ID CE170100009) and DP160100253. Authors also thank Prof. Jason Twamley for useful discussions. 

\section*{References}


%

\end{document}